\documentstyle[a4wide,12pt]{article}
\title{Coordinate transformations in Schwarzschild singularity problem}
\author{Vu B. Ho\\Department of Physics\\ Monash University\\ Clayton
Victoria 3168\\ Australia}
\date{23-11-93}
\begin{document}
\maketitle
\begin{abstract}
We consider a systematic approach to the coordinate transformations in
the Schwarzschild singularity problem in general relativity. It is shown
that not only the singularity at the surface $r=2m$ but also the
singularity at the origin $r=0$ can be eliminated by suitable coordinate
transformations.
\end{abstract}

\newpage
The Schwarzschild solution of the vacuum Einstein equations
$R_{\mu\nu}-1/2Rg_{\mu\nu}=0$ for static and spherically symmetric
spacetimes is normally written in the following form \cite{Mis}

\begin{equation}
ds^2=-\left(1-\frac{2m}{r}\right)dt^2+\left(1-\frac{2m}{r}\right)^{-1}dr^2
+ r^2\left(d\theta^2+sin^2\theta d\phi^2\right).
\end{equation}
With this form of solution, the metric components become singular at both
$r=2m$ and $r=0$. But it has been shown that while the singularity at $r=0$
is a genuine physical singularity, the singularity at $r=2m$ is only a
result of a breakdown of the coordinates \cite{Mis,Kru}. If we introduce
new coordinates $(X,T)$, Kruskal-Szekeres coordinates, according to the
following relations

\begin{eqnarray}
t&=&2m\ln\left(\frac{X+T}{X-T}\right),\\
\left(\frac{r}{2m}-1\right)\exp\left(\frac{r}{2m}\right)&=&(X+T)(X-T),
\end{eqnarray}
then the Schwarzschild metric takes the form

\begin{equation}
ds^2=\frac{32m^3}{r}\exp\left(-\frac{r}{2m}\right) \left(-dT^2+dX^2\right)
+r^2\left(d\theta^2+sin^2\theta d\phi^2\right),
\end{equation}
and the singularity at $r=2m$ is eliminated. The Kruskal-Szekeres coordinate
transformation can be derived using the outgoing and ingoing
radial null geodesics of the Schwarzschild spacetime \cite{Mis,Wald}.
In this paper we would like to discuss a method that will allow us to derive
in a more general way different good coordinates. For this purpose let us
introduce the following coordinate transformations

\begin{eqnarray}
t &=& g(X,T),\\
1-\frac{2m}{r} &=& f(X,T), \ \ \ \ \mbox{ or } \ \ \ \
r = \frac{2m}{1-f(X,T)},
\end{eqnarray}
where $g(X,T)$ and $f(X,T)$ are arbitrary functions that will be determined
later. By differentiation and substitution into the Schwarzschild metric (1)
we have

\begin{eqnarray}
-fdt^2 + \frac{1}{f}dr^2 &=& \left[-2f\frac{\partial g}{\partial T}
\frac{\partial g}{\partial X} + \frac{8m^2}{f(1-f)^4} \frac{\partial
f}{\partial T}\frac{\partial f}{\partial X}\right]dTdX\\
&+&\left[-f\left(\frac{\partial g}{\partial
T}\right)^2 + \frac{4m^2}{f(1-f)^4}\left(\frac{\partial f}{\partial
T}\right)^2\right]dT^2\\
&+& \left[-f\left(\frac{\partial g}{\partial X}
\right)^2 + \frac{4m^2}{f(1-f)^4}\left(\frac{\partial f}{\partial X}
\right)^2 \right]dX^2.
\end{eqnarray}
In order to put $ds^2$ in a form similar to (4), we require the functions
$f(X,T)$ and $g(X,T)$ satisfy the following system of equations

\begin{eqnarray}
\frac{\partial g}{\partial T}\frac{\partial g}{\partial X} &=& \left[
\frac{2m}{f(1-f)^2}\right]^2 \frac{\partial f}{\partial T} \frac{\partial
f}{\partial X}\\
\left(\frac{\partial g}{\partial T}\right)^2 + \left(\frac{\partial
g}{\partial X}\right)^2 &=& \left[\frac{2m}{f(1-f)^2}\right]^2 \left[\left(
\frac{\partial f}{\partial T}\right)^2 + \left(\frac{\partial f}{\partial
X}\right)^2\right].
\end{eqnarray}
{}From these relations we can derive the following simpler relations

\begin{eqnarray}
\frac{\partial g}{\partial X} + \frac{\partial g}{\partial T} &=& \pm
\frac{2m}{f(1-f)^2}\left(\frac{\partial f}{\partial X} + \frac{\partial f}
{\partial T}\right),\\
\frac{\partial g}{\partial X} - \frac{\partial g}{\partial T} &=& \pm
\frac{2m}{f(1-f)^2} \left(\frac{\partial f}{\partial X} - \frac{\partial
f}{\partial T}\right).
\end{eqnarray}
However we will restrict in this paper to coordinate transformations using
the following stronger conditions. It follows from the relations (10,11)
that we either require

\begin{eqnarray}
\frac{2m}{f(1-f)^2}\frac{\partial f}{\partial T}&=&
\pm\frac{\partial g}{\partial T},\\
\frac{2m}{f(1-f)^2}\frac{\partial f}{\partial X}&=&\pm\frac{\partial g}
{\partial X},
\end{eqnarray}
or require
\begin{eqnarray}
\frac{2m}{f(1-f)^2}\frac{\partial f}{\partial T}&=&
\pm\frac{\partial g}{\partial X},\\
\frac{2m}{f(1-f)^2} \frac{\partial f}{\partial X}&=&\pm\frac{\partial g}
{\partial T}.
\end{eqnarray}

Let us consider the conditions (16,17) first. In this case $ds^2$ can be
rewritten in a simpler form

\begin{equation}
ds^2 = f\left[\left(\frac{\partial g}{\partial T}\right)^2 - \left(
\frac{\partial g}{\partial X}\right)^2\right](-dT^2+dX^2) + r^2(d\theta^2 +
sin^2\theta d\phi^2).
\end{equation}
It is straightforward to verify that the Kruskal-Szekeres coordinate
transformations (2,3) satisfy these conditions, and it is also ready to show
that the Schwarzschild metric (4) can be derived from the metric form (18).
Now it is observed that the relations (16,17) can be put in the following
more convenient forms

\begin{eqnarray}
2m\frac{\partial}{\partial T}\left[\ln\left(\frac{f}{1-f}\exp\left(
\frac{1}{1-f}\right)\right)\right] &=& \pm\frac{\partial g}{\partial X},\\
2m\frac{\partial}{\partial X}\left[\ln\left(\frac{f}{1-f}\exp\left(
\frac{1}{1-f}\right)\right)\right] &=& \pm\frac{\partial g}{\partial T}.
\end{eqnarray}
It is clear from these forms if we can find a function $G(X,T)$ so that the
below conditions are satisfied

\begin{equation}
\frac{\partial g}{\partial X} = \frac{\partial G}{\partial T}, \ \ \ \ \ \
\frac{\partial g}{\partial T} = \frac{\partial G}{\partial X},
\end{equation}
then we immediately obtain a general form of coordinate transformations
for the variable $r$ as follows

\begin{equation}
\frac{f}{1-f}\exp\left(\frac{1}{1-f}\right)=\exp\left(\pm\frac{G}{2m}\right).
\end{equation}
On the other hand, from (21) we also obtain a wave condition for the
function $g(X,T)$

\begin{equation}
\frac{\partial^2 g}{\partial X^2}-\frac{\partial^2 g}{\partial T^2}=0.
\end{equation}
So with respect to the new coordinates $(X,T)$ the time
$t$ of the coordinates $(r,t)$ behaves like a wave, and its relationship
with the new coordinates $(X,T)$ can be put in a general form

\begin{equation}
t=g(X,T) = g_L(X+T)+ g_R(X-T),
\end{equation}
where $g_L$ and $g_R$ are arbitrary functions of the coordinates $(X,T)$. The
transformation (2) is a particular solution of the wave equation (23) since
it can be rewritten in the form $g(X,T)=2m\ln(X+T)-2m\ln(X-T)$.

As another illustration let us consider the simple case when $f$ depends
only on $X$ while $g$ depends only on $T$. In this case from (10,11) we
have the following simple relation

\begin{equation}
\frac{2m}{f(1-f)^2} \frac{\partial f}{\partial X} = \pm\frac{\partial
g}{\partial T} = C,
\end{equation}
where $C$ is some constant. Integrating this system we find

\begin{equation}
g = \pm CT, \ \ \ \ \ \
\frac{f}{(1-f)}\exp\left(\frac{1}{1-f}\right) =
\exp\left(\frac{CX}{2m}\right).
\end{equation}
In terms of $r$ and $t$ we have

\begin{equation}
g = \pm CT, \ \ \ \ \ \
\left(\frac{r}{2m}-1\right) \exp\left(\frac{r}{2m}\right) =
\exp\left(\frac{CX}{2m}\right).
\end{equation}
In this case the Schwarzschild metric takes a simple form

\begin{equation}
ds^2 = C^2(\frac{r-2m}{r})(-dT^2+dX^2)+r^2(d\theta^2+sin^2\theta d\phi^2).
\end{equation}
It is seen from (27) that the new coordinates $(X,T)$ correspond to the
region $r>2m$ only. The surface $r=2m$ is seen as infinity of $X$, whose
sign depends on the sign of the constant of integration $C$. The new time
$T$ may flow forward as the old time $t$ or flow backward with respect to
$t$. In a similar manner we can consider the case when $f$ is a function of
$T$ only and $g$ of $X$ only. In this case we obtain

\begin{equation}
t=\pm CX, \ \ \ \ \ \
\left(\frac{r}{2m}-1\right) \exp\left(\frac{r}{2m}\right) =
\exp\left(\frac{C_1T}{2m}\right).
\end{equation}
Now in the new coordinates $(X,T)$, time only begins $(T=-\infty)$ or ends
$(T=+\infty)$ at the surface $r=2m$, depending on the sign of the constant
of integration $C$. There is no real time for the region $r<2m$.

Now let us consider the conditions (14,15). In this case if we put the
function $g(X,T)$ in the form

\begin{equation}
g(X,T) = 2m\ln F(X,T),
\end{equation}
where $F$ is an arbitrary function of $(X,T)$, then we obtain immediately
the following form of coordinate transformations for the coordinate $r$

\begin{equation}
\left(\frac{r}{2m}-1\right) \exp\left(\frac{r}{2m}\right) = F(X,T),
\ \ \ \ (\mbox{ or } \frac{1}{F(X,T)}).
\end{equation}
With the Kruskal-Szekeres time coordinate transformation as in (2), we have
$F(X,T)=(X+T)/(X-T)$ and the coordinate transformation for $r$ in this case
takes the form

\begin{equation}
\left(\frac{r}{2m}-1\right) \exp\left(\frac{r}{2m}\right)=\frac{X+T}{X-T}.
\end{equation}
In this form we see that the origin $r=0$ corresponds to the time axis
$X=0$ in the $(X,T)$ coordinates, while the surface $r=2m$ is transformed
into the line $X=-T$.

When the conditions (14,15) are satisfied the Schwarzschild metric reduces to
that of a spherical surface in three dimensional Euclidean space

\begin{equation}
ds^2 = r^2(d\theta^2+sin^2\theta d\phi^2).
\end{equation}
This is a remarkable result, since we see that now not only the singularity
at $r=2m$ but also the singularity at $r=0$ is eliminated. This interesting
feature may be seen from the type of coordinate transformations, because
we can rewrite (14,15) in a form that gives a direct relationship between the
coordinates $r$ and $t$ as follows

\begin{equation}
\left(\frac{r}{2m}-1\right)\exp\left(\frac{r}{2m}\right) =
\exp\left(\pm\frac{t}{2m}\right).
\end{equation}
We see that the time $t$ either begins or ends at the surface $r=2m$. For
the region $r<2m$, time does not exist unless it is complex and of
the form $2m(a+(2n+1)\pi i)$, where $a$ is a real number and $n$ an integer.
The rates of the variable $r$ can be found by differentiation of the
equation (34). With the positive sign before $t$ in (34), we find

\begin{equation}
\frac{dr}{dt}=\left(1-\frac{2m}{r}\right), \ \ \ \ \ \ \frac{d^2r}{dt^2} =
\frac{2m}{r^2}\left(1-\frac{2m}{r}\right).
\end{equation}
In this case space is expanding with an increasing rate until the rate
reaches the speed of light at infinity. Time begins at the surface $r=2m$.
With the negative sign we obtain

\begin{equation}
\frac{dr}{dt}=-\left(1-\frac{2m}{r}\right), \ \ \ \ \ \ \frac{d^2r}{dt^2} =
\frac{2m}{r^2}\left(1-\frac{2m}{r}\right).
\end{equation}
Now space is contracting until the surface $r=2m$ is reached and time will
end there. If we identify this physical process with some sort of gravity,
then gravitation may be considered as a manifestation of space contraction.

\section*{Acknowledgements}
I would like to acknowledge the financial support of APA Research Award.

\end{document}